\documentclass[10pt,letterpaper]{article}

\usepackage{amsmath}
\usepackage{latexsym,amssymb}
\usepackage{opex3}
\usepackage{cite}

\begin{document}
\pagestyle{plain}
\setlength{\topmargin}{39pt}


\title{Electromagnetic Lorenz Fields}

\author{H. C. Potter}

\address{Behavior Theory Institute, Ste.~110, 1122 Garvin Pl., \\Louisville, KY 40203, USA}

\email{hcp40143@gmail.com} 




\begin{abstract*}
Gauge transformations are potential transformations that leave only
specific Max\-well fields invariant. To reveal more, I develop Lorenz field equations
with full Maxwell form for nongauge, sans gauge function, transformations
yielding mixed, superposed retarded and outgoing, potentials. The
form invariant Lorenz condition is then a charge conservation equivalent.
This allows me to define three transformation classes that screen
for Lorenz relevance. The nongauge Lorentz conditions add polarization
fields which support emergent, light-like rays that convey energy
on charge conserving phase points. These localized rays escape discovery
in modern Maxwell fields where the polarizations are suppressed by
gauge transformations.
\end{abstract*}





\section{Introduction}

In 1867, during the time when J. C. Maxwell (1831-79) was publishing
his electromagnetic theory, L. V. Lorenz (1829-91) published his theory
equating light vibrations with electric currents \cite{Lorenz}. This
work is translated to modern vector notation and critiqued in \cite{Potter}.
Starting from Kirchhoff's Ohm's law expression, Lorenz uses scalar
potential retardation to obtain an expression for the retarded vector
potential. The retarded potentials satisfy inhomogeneous wave equations
with sources and, when charge is conserved, the eponic Lorenz condition
\cite[pp. 268-9]{Whittaker}. Because the retarded potentials are
nonlocal, all their derivatives also must be inhomogeneous wave function
solutions for correspondingly differentiated remote sources. Since
the potentials satisfy wave equations they can be augmented with solutions
for the homogeneous wave equation. These outgoing augmentations are
local, their derivatives depend only on proximate values. Although
no longer strictly retarded, the composite potentials still satisfy
a wave equation with one and the same propagation speed, but the Lorenz
condition is modified.

In his 1867 paper, Lorenz never mentions the magnetic field and, therefore,
never develops electromagnetic field equations. In fact, until now,
this extension has not been examined. However, by defining magnetic
induction as the vector potential curl, I derive electromagnetic Lorenz
field equations with the Maxwell form. This derivation allows Lorenz
retardation ramifications to be fully explored. When derived from
nonlocal potentials, the fields in these equations must be nonlocal
also. The analytically desirable locality can be restored as a far
field approximation for systems that satisfy the dipole approximation
which imposes specific size constraints \cite[p. 222]{Marion}. For
the augmented potentials, the Lorenz field equations incorporate fields
for electric displacement and magnetic field strength by adding local
 polarizations. To help understand the effect Lorenz retardation can
have on light theory, this development is presented formally in Appendix
\ref{sec:Lorenz-Field-Equatons}. It shows that wave function potentials
satisfying the Lorenz condition assure charge conservation. Thus,
the form invariant Lorenz condition and charge conservation are equivalent.

\section{Potentials transformation}

In the early twentieth century, starting from electromagnetic Maxwell
field equations, H. A. Lorentz (1853-1928) found a condition that
caused the vector and scalar potentials to be wave functions with
the retarded  forms. This Lorentz condition is the same mathematical
equation found earlier by Lorenz. Since the Lorenz condition is equivalent
to charge conservation, Maxwell fields must always be derivable from
wave function potentials.

Having started from the field equations, Lorentz noted that the electric
field strength and magnetic induction are invariant when the potentials
are subjected to transformation with an arbitrary gauge function,
$\chi$, that causes\begin{equation}
\left.\begin{array}{ccc}
\mathbf{A} & = & \mathbf{A}_{1}-\boldsymbol{\nabla}\chi,\\
\Omega & = & \Omega_{1}+\frac{1}{a}\frac{\partial\chi}{\partial t}.\end{array}\right\} \label{eq:GaugeTransform}\end{equation}
Gauge restraints on potentials are now widely accepted in theoretical
physics \cite{JacksonOkun} even without gauge function specification
\cite{Heras} which further restrains the allowed potentials \cite{Jackson}.
Although nonlocality and susceptibility to gauge restraint causes
potential reality to be questioned, arbitrary gauge restraint permissibility
is unquestioned. However, charge conservation equivalence to the Lorenz
condition exposes this deficiency and imposes a requirement for extraordinary
justification on any transformation that alters the Lorenz condition. 

When the original potentials are retarded wave functions Eq.(\ref{eq:GaugeTransform})
resolves to three Lorenz classes that need not have the gauge transformation
form. First, the new potentials are not  wave functions. Second, the
new potentials are  wave functions with altered charge and current
values. Third, the new potentials are  wave functions with unaltered
charge and current values. In the first case the Lorentz condition
will not apply and the potentials will not be retarded; whereas recent
tests listed in \cite{Retardation1,Retardation2,Retardation3,Retardation4}
show longitudinal electric fields to propagate with finite speed.
The last two cases will conserve charge if  the Lorentz condition
persists. This persistence will assure Lorenz field equations in which
charge and current densities are deemed observables that produce physical
potentials and fields yielding emergent light-like waves that conserve
charge progressively as described in Appendix \ref{sec:Polarization-Waves}
with localization to ray forming phase points on which energy is conveyed
as described in Appendix \ref{sec:Polarization-Flux-Localization}.
For gauge transformations, this requires the Lorenz gauge, $\square_{a}\chi=0$,
that suppresses local electric and magnetic polarizations and, thus,
their dependent, emergent rays that have been fruitlessly sought in
Maxwell fields \cite{Keller}.

\section{Conclusion}

The forgoing shows that if Lorenz had developed field equations from
his retarded potentials classical electromagnetism as we know it today
could have been provided with retarded potentials as a solid etiological
foundation. We would fully appreciate the Lorenz condition equivalence
to charge conservation, light-like ray emergence from retarded fields
and energy conveyance by field phase points. Based on this analysis
I propose that distinct terms {}``Lorenz condition'' and {}``Lorentz
condition'' used in the Abstract be retained to designate  two uses
for one equation relating vector and scalar potentials: the Lorentz
condition only representing a potential transformation; the Lorenz
condition representing charge conservation that potential transformations
leave form invariant, Lorenz covariance. With this, Lorenz fields
support localized, light-like rays that escape discovery in modern
Maxwell fields. 

\appendix

\section{Lorenz Field Equations\label{sec:Lorenz-Field-Equatons}}

The electromagnetic Lorenz potentials are given by the  forms

\begin{equation}
\Omega=\int\int\int\frac{\mathbf{\mathrm{\rho'}}}{R}dv'\label{eq:Omega}\end{equation}
and

\begin{equation}
\mathbf{A}=\frac{1}{a}\int\int\int\frac{\mathbf{J}'}{R}dv'.\label{eq:A}\end{equation}
where $\mathbf{\mathrm{\rho'}}$ and $\mathbf{\mathbf{J'}}$ indicate
that $\mathbf{\mathrm{\rho}}$ and $\mathbf{\mathbf{J}}$ depend on
$t-\frac{R}{a}$ and $\mathbf{r}'$ rather than $t$ and $\mathbf{r}$
for $\mathbf{R}=\mathbf{r}-\mathbf{r}'$. They satisfy the wave equations
\begin{equation}
\square_{a}\Omega=-4\pi\rho\label{eq:OmegaWave}\end{equation}
and\begin{equation}
\square_{a}\mathbf{A}=-\frac{4\pi}{a}\mathbf{\mathbf{J}}\label{eq:Awave}\end{equation}
 and, when charge is conserved,  the Lorenz condition\begin{equation}
-a\boldsymbol{\nabla}\bullet\mathbf{A}=\frac{\partial\Omega}{\partial t}.\label{eq:RetardedLorenzCondition}\end{equation}
Here, the D'Alembertian, $\square_{a}=\frac{\partial^{2}}{\partial x^{2}}+\frac{\partial^{2}}{\partial y^{2}}+\frac{\partial^{2}}{\partial z^{2}}-\frac{1}{a^{2}}\frac{\partial^{2}}{\partial t{}^{2}}$.
When the potentials are augmented by solutions to the homogeneous
wave equation, $\square_{a}\Omega_{0}=\square_{a}\mathbf{A}_{0}=0$,
the Lorenz condition takes the modified form\begin{equation}
-a\boldsymbol{\nabla}\bullet(\mathbf{A}+\mathbf{A}_{0})=\frac{\partial(\Omega+\Omega_{0})}{\partial t}.\label{eq:AugmentedRetardedLorenzCondition}\end{equation}
The augmentations contain implicit scalar strengths since one or both
could be set to zero. For\begin{equation}
\mathbf{E}=-\boldsymbol{\nabla}\Omega-\frac{1}{a}\frac{\partial\mathbf{A}}{\partial t}\label{eq:E}\end{equation}
and\begin{equation}
\mathbf{\mathbf{B}}=\mathbf{\boldsymbol{\nabla}}\times\mathbf{A},\label{eq:B}\end{equation}
we have the additional fields $\mathbf{D}=\mathbf{E}+\mathbf{P}$
and $\mathbf{\mathbf{H}}=\mathbf{\mathbf{B}}-\mathbf{M}$ with $\mathbf{P}=-\nabla\Omega_{0}-\frac{1}{a}\frac{\partial\mathbf{A}_{0}}{\partial t}$
and $\mathbf{M}=-\mathbf{\boldsymbol{\nabla}}\times\mathbf{A}_{0}$.
In these definitions the augmentation strengths are passed to the
polarizations which vanish when the augmentations are derived from
a gauge function, \emph{i.e.} when $\mathbf{A}_{0}=-\boldsymbol{\nabla}\chi$
and $\Omega_{0}=\frac{1}{a}\frac{\partial\chi}{\partial t}$. These
fields satisfy the equations:\begin{subequations}\label{L:joint}\begin{equation}
\boldsymbol{\nabla}\bullet\mathbf{\mathbf{B}}=0,\label{eq:L:B}\end{equation}
\begin{equation}
\boldsymbol{\nabla}\bullet\mathbf{D}=4\pi\rho,\label{eq:L:E}\end{equation}
\begin{equation}
a\mathbf{\boldsymbol{\nabla}}\times\mathbf{E}=-\frac{\partial\mathbf{\mathbf{B}}}{\partial t},\label{eq:L:F}\end{equation}
\begin{equation}
\mathbf{\boldsymbol{\nabla}}\times\mathbf{H}=\frac{1}{a}\left[\frac{\partial\mathbf{D}}{\partial t}+4\pi\mathbf{J}\right].\label{eq:L:A}\end{equation}
\end{subequations}Together Eqs.(\ref{eq:L:E}) and (\ref{eq:L:A})
give the condition for charge conservation\begin{equation}
\boldsymbol{\nabla}\bullet\mathbf{\mathbf{J}}+\frac{\partial\rho}{\partial t}=0.\label{eq:ChargeContinuity}\end{equation}
This also follows from simply applying the D'Alembertian to the Eq.(\ref{eq:AugmentedRetardedLorenzCondition})
Lorenz condition.  When $\mathbf{E}$ and $\mathbf{B}$ in the field
equations are replaced by $\mathbf{P}$ and $\mathbf{M}$ using the
expressions for $\mathbf{D}$ and $\mathbf{H}$, from $\mathbf{H}\bullet(\boldsymbol{\nabla}\times\mathbf{\mathbf{D}})-\mathbf{D}\bullet(\boldsymbol{\nabla}\times\mathbf{\mathbf{H}})=\boldsymbol{\nabla}\bullet(\mathbf{\mathbf{D}}\times\mathbf{H})$
Eqs.(\ref{eq:L:F}) and (\ref{eq:L:A}) give for energy continuity\begin{equation}
\boldsymbol{\nabla}\bullet(\mathbf{\mathbf{D}}\times\mathbf{H})=-\frac{1}{a}\left[\mathbf{H}\bullet\frac{\partial\mathbf{H}}{\partial t}+\mathbf{D}\bullet\frac{\partial\mathbf{D}}{\partial t}\right]-\frac{4\pi}{a}\mathbf{\mathbf{D}}\bullet\mathbf{\mathbf{J}}.\label{eq:DxHcontinuity}\end{equation}
More directly,\begin{equation}
\boldsymbol{\nabla}\bullet(\mathbf{\mathbf{E}}\times\mathbf{H})=-\frac{1}{a}\left[\mathbf{H}\bullet\frac{\partial\mathbf{B}}{\partial t}+\mathbf{E}\bullet\frac{\partial\mathbf{D}}{\partial t}\right]-\frac{4\pi}{a}\mathbf{E}\bullet\mathbf{\mathbf{J}}.\label{eq:ExHcontinuity}\end{equation}
These expressions are identical when the polarizations vanish, but
when $\mathbf{E}$ and $\mathbf{B}$ vanish the latter vanishes and
the former carries the entire flux. This coupling allows local polarization
waves to emerge from a nonlocal electromagnetic field. The Hertz dipole
radiation solution discussed in Appendices \ref{sec:Polarization-Waves}
and \ref{sec:Polarization-Flux-Localization} is the only extant example
for which this emergence is actually exhibited.

For Lorenz fields the coupling normally provided by the constitutive
equations, $\mathbf{D}=\epsilon\mathbf{E}$ and $\mathbf{B}=\mu\mathbf{H}$,
can not be invoked without further augmenting the potentials with
proportional retarded potentials, because fields derived from
the retarded potentials can not be proportional to those derived from
the potential augmentations. So the wave function speed must be adjusted
for dielectric constant $\epsilon$ and magnetic permeability $\mu$
by measurement. Being dependent on retarded potentials, all fields
satisfy wave equations with the same propagation speed. Augmented potential 
polarization waves satisfying the Lorenz condition are
discussed in Appendix \ref{sec:Polarization-Waves}.

\section{Polarization Waves\label{sec:Polarization-Waves}}

As representative Lorenz potential augmentations consider\begin{equation}
\mathbf{A}_{0}=\left(f,g,h\right)\label{eq:Azero}\end{equation}
and\begin{equation}
\Omega_{0}=\frac{\mathbf{k}}{k}\bullet\mathbf{A}_{0}=\frac{k_{x}f+k_{y}g+k_{z}h}{k}.\label{eq:OmegaZero}\end{equation}
When $f$, $g$ and $h$ are functionally dependent only on phase
factors $\omega t-\mathbf{k}\bullet\mathbf{r}$ where $\mathbf{k}=(k_{x},k_{y},k_{z})$
and $a^{2}k^{2}=\omega^{2}$, these potentials satisfy the homogeneous
wave equation. Potentials in matter free space have not been restrained
previously by the Eq.(\ref{eq:RetardedLorenzCondition}) Lorenz condition
as these are. With this restraint, the potentials describe waves that
can be considered to propagate by progressively conserving charge.
An example with Lissajous vector potential is presented in Appendix
\ref{sec:Polarization-Flux-Localization}. It shows that the polarizations
will describe rays without the longitudinal fields that plague other
formulations \cite{Keller}. The rays are real space paths determined
by three linearly independent potential vector component phases. With
neither wave packet dispersion \cite{Darwin} nor quantum mechanical
nonlocality, these rays provide localization that has long eluded
discovery  \cite{Keller}. Localization to rays is relaxed when
the component phases are linearly dependent.

For the Eq.(\ref{eq:Azero}) and (\ref{eq:OmegaZero}) potentials,
the electric and magnetic polarizations become\begin{equation}
\mathbf{P}=\frac{1}{\omega}\frac{\mathbf{k}}{k}\times(\mathbf{k}\times\dot{\mathbf{A}_{0}})=
\frac{1}{\omega}(\frac{\mathbf{k}}{k}\bullet\dot{\mathbf{A}_{0}})\mathbf{k}-
\frac{1}{\omega}(\frac{\mathbf{k}}{k}\bullet\mathbf{k})\dot{\mathbf{A}_{0}},\label{eq:P}\end{equation}
since $\boldsymbol{\nabla}\Omega_{0}=-\frac{\mathbf{k}}{\omega}\dot{\Omega_{0}}$,
and \begin{equation}
\mathbf{M}=\frac{1}{\omega}(\mathbf{k}\times\dot{\mathbf{A}_{0}}).\label{eq:M}\end{equation}
When the vector potential components are periodic functions they can
be considered to be plane waves. Unlike Lorenz fields defined by nonlocal
potentials, the polarizations should be considered to be local point
functions. They are orthogonal, $\mathbf{P}\bullet\mathbf{M}=0$ and
have no longitudinal components, $\mathbf{k}$$\bullet$$\mathbf{P}$=$\mathbf{k}$$\bullet$$\mathbf{M}$=0;
they carry a flux $\mathbf{M}\times\mathbf{P}=\omega^{-2}(\mathbf{k}\times\dot{\mathbf{A}_{0}})^{2}\frac{\mathbf{k}}{k}$
equal to $\mathbf{D}\times\mathbf{H}$ when $\mathbf{E}=\mathbf{B}=0$
and travel indefinitely in a ray direction $\mathbf{k}$ at speed
$a$ without driving sources.

In the 19$^{th}$ century second half, H. von Helmholtz (1821-94)
attempted to reconcile competing electromagnetic theories \cite{Woodruff}.
To this end he developed a theory based on electrical and magnetic
polarization and obtained wave equations for polarization propagation
in a homogeneous medium. His wave equation for the electric polarization
contained an undetermined constant that allowed the propagation speed
for longitudinal waves to have any non-negative value. So only difficult,
precision measurements to establish the parameter's value could complete
the theory. For isotropic electric polarization his theory gives $c/\sqrt{\epsilon\mu}$
for component propagation speed, where $c$ is the vacuum light speed.
The Lorenz electric polarization does propagate isotropically. But
we have just seen that the Lorenz condition suppresses its longitudinal
component, because Eq.(\ref{eq:L:E}) gives $\boldsymbol{\nabla}\bullet\mathbf{\mathbf{P}}=0$.

To promote his theory, Helmholtz proposed a prize competition to experimentally
establish a relation between electromagnetism and dielectric polarization.
His former student, H. Hertz (1857-94), later claimed the prize  and
went on to observe electromagnetic reflection and interference. Based
on these observations Hertz concluded that polarization propagation
is analogous to vacuum light \cite[pp. 19 and 122-3]{Hertz} and \cite{Hertz2}.
 He studied  waves in a hall with effective meter dimensions 15x8.5x6.
These waves emanated from a 1 cm spark gap with calculated 14 ns resonant
half period. Powerful discharges were obtained by applying interrupted
induction coil output across the gap between two 15 cm radius spheres
in a 100 cm long dipole with 40 cm by 40 cm brass plates at its far
ends. In explaining his results, Hertz applied Maxwell's equations
to the cyclic gap charging and neglected the periodic arc discharge
\cite{Marion,Hertz,Hertz2}. Today, we should prefer an alternative
description in which the arc powered by the collapsing scalar potential
is the radiation cause \cite{Lindell,Lindell2,Lindell3}, because
the arc  generates light and induces remote circuit arcing that Hertz
used to analyze the  wave fields. Apparently arcing at his remote
circuits was stroboscopically sensitized by light from his primary
arc, so he could use only primary spark gaps near 1 cm where light
emission was sufficient to make observations. Thus, Hertz's dipole
radiation fields displayed in text books are unlikely to represent
the fields he actually studied; but, if photons have finite dimensions,
his and recent \cite{PhaseSkip} observations suggest their internal
structure may be susceptible to study. The discussion in \cite{Potter} 
shows that observations documented in \cite{PhaseSkip} may represent 
an Eq.(\ref{eq:L:A}) Ampere law based, magnetoinductive internal energy 
structure with anti-phased electric and magnetic fields.

\section{Polarization Flux Localization\label{sec:Polarization-Flux-Localization}}

For vacuum polarization waves Eq.(\ref{eq:DxHcontinuity})  takes
the form\begin{equation}
\boldsymbol{\nabla}\bullet(\mathbf{\mathbf{M}}\times\mathbf{P})=
-\frac{1}{a}\left[\mathbf{M}\bullet\frac{\partial\mathbf{M}}
{\partial t}+\mathbf{P}\bullet\frac{\partial\mathbf{P}}{\partial t}
\right].\label{eq:VacuumEnergyContinuity}\end{equation}
By the Gauss theorem, this represents an equality between the temporal
energy change in a volume and the energy flux through its surface.
For monochromatic waves, $\mathbf{\mathbf{M}}\times\mathbf{P}$ can
be written using the Hertz analogy as $h\nu\mathbf{f}$ where $h\nu$
is photon energy and $\mathbf{f}$ is photon flux with coherence length
inversely related to monochromaticity departure \cite{Mandel}. This
dependence on chromaticity means that the photon power density must
approach zero as the coherence length becomes very large for monochromatic
photons. So the compact , quantum particle photon concept is untenable
in this limit. The concept is further threaten by confounding photon
size with wavelength. This problem is revealed by the sensible benchmarks
for photon properties ranging over 10$^{30}$ ev in Table \ref{Table:PhotonCharacteristics}
where the lowest energy will have a wavelength greater than the Earth
orbit radius.

\begin{table}[h]

\caption {\large Some Vacuum Photon Properties\label{Table:PhotonCharacteristics}}

{\small \noindent \begin{raggedright}{Electron rest mass equals 0.5 Mev, visible
light at 5000$\textrm{\AA}$ equals 248ev, and Cosmic Microwave Background
at 3 K equals 0.26 mev. Atomic nucleus radii are about E-12 cm, atomic
radii are about E-8 cm, Earth radius is 6.38E8 cm and Earth orbit
radius is 1.5E13 cm. Solar radiant energy flux at Earth is 8.58E21
ev/(s m$^{2}$) with an energy distribution that should be appropriate
for the 5780 K effective sun temperature. The references indicate
internal structure studies.}
\bigskip{}
\par\end{raggedright}} \normalsize

\begin{centering}{\LARGE }\begin{tabular}{ccc}
\hline
Energy&
Frequency &
Wavelength \tabularnewline
{\small $h\nu$}&
{\small $\nu$ Hz }&
{\small $\lambda$ cm }\tabularnewline
\hline
\hline 
1Tev &
0.241E27 &
12.4E-17 \tabularnewline
1Gev &
0.241E24&
12.4E-14\tabularnewline
1Mev &
0.241E21&
12.4E-11\tabularnewline
1kev &
0.241E18&
12.4E-8\tabularnewline
1ev &
0.241E15&
12.4E-5\tabularnewline
1mev&
 0.241E12&
12.4E-2\tabularnewline
1$\mu$ev \cite{Hertz}&
0.241E9&
12.4E1\tabularnewline
1nev &
0.241E6 &
12.4E4\tabularnewline
1pev &
0.241E3&
12.4E7\tabularnewline
1fev&
0.241&
12.4E10\tabularnewline
1aev \cite{PhaseSkip}&
0.241E-3&
12.4E13\tabularnewline
\hline
\end{tabular}\par\end{centering}

\end{table}

As mentioned in Appendix \ref{sec:Polarization-Waves}, the Lorenz
field polarization waves provide, at least, a classical solution to
this conceptual impasse. To see how, consider the vector potential\begin{equation}
\mathbf{A}_{0}=\left(\sin\phi_{1},\sin\phi_{2},\sin\phi_{3}\right)\label{eq:ExAzero}\end{equation}
with Lissajous phases $\phi_{1}=\omega_{1}t-\mathbf{k}_{1}\bullet\mathbf{r}$,
$\phi_{2}=\omega_{2}t-\mathbf{k}_{2}\bullet\mathbf{r}$ and $\phi_{3}=\omega_{3}t-\mathbf{k}_{3}\bullet\mathbf{r}$
where $\mathbf{k}_{1}=(k_{1x},k_{1y},0)$, $\mathbf{k}_{2}=(k_{2x},k_{2y},0)$
and $\mathbf{k}_{3}=(0,0,k_{3z})$. This vector potential may be looked
upon as originating from a charge moving in an x-y plane. When $a^{2}k_{I}^{2}=\omega_{I}^{2}$
for $I=1,2\textrm{ or }3$, $\square_{a}\mathbf{A}_{0}=0$. To satisfy
the Eq.(\ref{eq:RetardedLorenzCondition}) Lorenz condition, we find
the scalar potential must have the form \begin{equation}
\Omega_{0}=\frac{k_{1x}}{k_{1}}\sin\phi_{1}+\frac{k_{2y}}{k_{2}}\sin\phi_{2}+\sin\phi_{3}\label{eq:ExOmegaZero}\end{equation}
which also satisfies $\square_{a}\Omega_{0}=0$. So these potentials
can be taken to define the Appendix \ref{sec:Lorenz-Field-Equatons}
Lorenz polarizations given by\begin{equation}
\mathbf{P}=(\frac{k_{2x}k_{2y}}{k_{2}}\cos\phi_{2}-\frac{k_{1y}^{2}}{k_{1}}\cos\phi_{1},
\frac{k_{1x}k_{1y}}{k_{1}}\cos\phi_{1}-\frac{k_{2x}^{2}}{k_{2}}\cos\phi_{2},0)\label{eq:ExP}\end{equation}
and \begin{equation}
\mathbf{M}=(0,0,k_{2x}\cos\phi_{2}-k_{1y}\cos\phi_{1})\label{eq:ExM}\end{equation}
with the flux\begin{equation}\begin{split}
\mathbf{M}\times\mathbf{P}=(k_{2x}\cos\phi_{2}&-k_{1y}\cos\phi_{1})\times\\
&\times(-\frac{k_{1x}k_{1y}}{k_{1}}\cos\phi_{1}+\frac{k_{2x}^{2}}{k_{2}}\cos\phi_{2},
\frac{k_{2x}k_{2y}}{k_{2}}\cos\phi_{2}-\frac{k_{1y}^{2}}{k_{1}}\cos\phi_{1},0).
\end{split}\label{eq:ExFlux}\end{equation}
Taking $\phi_{1}=\phi_{2}=\phi$ gives the simple harmonic polarizations\begin{equation}
\mathbf{P}=\cos\phi(\frac{k_{2x}k_{2y}}{k_{2}}-\frac{k_{1y}^{2}}{k_{1}},
\frac{k_{1x}k_{1y}}{k_{1}}-\frac{k_{2x}^{2}}{k_{2}},0)\label{eq:ExShP}\end{equation}
and \begin{equation}
\mathbf{M}=\cos\phi(0,0,k_{2x}-k_{1y})\label{eq:ExShM}\end{equation}
with the flux\begin{equation}
\mathbf{M}\times\mathbf{P}=(k_{2x}-k_{1y})\cos^{2}\phi(-\frac{k_{1x}k_{1y}}{k_{1}}+\frac{k_{2x}^{2}}{k_{2}},
\frac{k_{2x}k_{2y}}{k_{2}}-\frac{k_{1y}^{2}}{k_{1}},0)\label{eq:ExShFlux}\end{equation}
having bead-chain \cite{Piekara} squared magnitude\begin{equation}
\left|\mathbf{M}\times\mathbf{P}\right|^{2}=(k_{2x}-k_{1y})^{2}[k_{2x}^{2}-2\frac{k_{2x}k_{1y}}{k_{1}k_{2}}
(k_{2x}k_{1x}+k_{2y}k_{1y})+k_{1y}^{2}]\cos^{4}\phi.\label{eq:ExShFlux^2}\end{equation}

When $\mathbf{k}_{1}=\mathbf{k}_{2}=\mathbf{k}/\sqrt{2}=(k_{x},k_{y},0)/\sqrt{2}$,
the polarizations further simplify to \begin{equation}
\mathbf{P}=\frac{1}{\sqrt{2}}\frac{k_{x}-k_{y}}{k}\cos\phi(k_{y},-k_{x},0)\label{eq:ExSmShP}\end{equation}
and \begin{equation}
\mathbf{M}=\frac{1}{\sqrt{2}}\cos\phi(0,0,k_{x}-k_{y})\label{eq:ExSmShM}\end{equation}
with the flux in the $\mathbf{k}$ direction and having squared magnitude\begin{equation}
\left|\mathbf{M}\times\mathbf{P}\right|^{2}=\frac{1}{4}(k_{x}-k_{y})^{4}\cos^{4}\phi.\label{eq:ExSmShFlux^2}\end{equation}
This flux is highly anisotropic with maximum values for $k_{x}=-k_{y}$
and null values for $k_{x}=k_{y}$.

When $\mathbf{k}_{1}\neq\mathbf{k}_{2}$ the phase moves on the parametric
line with coordinates\begin{equation}
\begin{array}{ccc}
X(t) & = & \left[k_{2y}(\omega_{1}t-\phi)-k_{1y}(\omega_{2}t-\phi)\right]/(k_{1x}k_{2y}-k_{2x}k_{1y}),\\
Y(t) & = & \left[-k_{2x}(\omega_{1}t-\phi)+k_{1x}(\omega_{2}t-\phi)\right]/(k_{1x}k_{2y}-k_{2x}k_{1y})\end{array}\label{eq:ExShCoords}\end{equation}
and velocity\begin{equation}
\mathbf{V}=(k_{2y}\omega_{1}-k_{1y}\omega_{2},k_{1x}\omega_{2}-k_{2x}\omega_{1},0)/(k_{1x}k_{2y}-k_{2x}k_{1y})\label{eq:ExShV}\end{equation}
with squared magnitude $V^{2}=2a^{2}k_{1}k_{2}\left[k_{1}k_{2}-(k_{1x}k_{2x}+k_{1y}k_{2y})\right]/(k_{1x}k_{2y}-k_{2x}k_{1y})^{2}$. If the flux vector
were to have a component perpendicular to the phase velocity the ray
would dissipate. So the vector product components must vanish or,
for the Eqs.(\ref{eq:ExShFlux}) and (\ref{eq:ExShV}) flux and
velocity,\begin{equation}
(k_{2y}\omega_{1}-k_{1y}\omega_{2})(\frac{k_{2x}k_{2y}}{k_{2}}-\frac{k_{1y}^{2}}{k_{1}})-(k_{1x}\omega_{2}-k_{2x}\omega_{1})(-\frac{k_{1x}k_{1y}}{k_{1}}+\frac{k_{2x}^{2}}{k_{2}})=0.\label{eq:ExShFxV}\end{equation}
This reduces to \begin{equation}
(k_{1y}+k_{2x})\left[k_{1}k_{2}-(k_{1x}k_{2x}+k_{2y}k_{1y})\right]=0\label{eq:ExShCond}\end{equation}
which gives two alternative conditions to be satisfied by the $\mathbf{k}_{1}$
and $\mathbf{k}_{2}$ components for nontrivial flux direction alignment
with the phase velocity direction. The phase $\varphi$ is a parameter
that gives simple harmonic polarizations. But $\varphi$ is not a
wave phase factor, because it is confined to the line defined by Eqs.(\ref{eq:ExShCoords}).
In the limiting case for which $\mathbf{k}_{1}=\mathbf{k}_{2}=\mathbf{k}/\sqrt{2}=(k_{x},k_{y},0)/\sqrt{2}$,
this ray-like character can be taken to persist even though this limit
also allows the conventional plane wave phase characterization for
$\varphi$. Failure to recognize this distinction as a real, physical
possibility has prevented these light-like rays from being extracted
from electromagnetic fields to help describe ray-like behavior in
geometrical optics and photography \cite{Duality1,Duality2}.

Although the case for $\mathbf{k}_{1}\neq\mathbf{k}_{2}$would appear
to have physical relevance in describing light-like waves, the more
general case in which $\varphi_{1}\neq\varphi_{2}$ may provide a
particle description for light. In this case, the phase point $(\varphi_{1},\varphi_{2})$
moves on the parametric path defined by equations like those in Eq.(\ref{eq:ExShCoords})
with $\varphi$ replaced by $\varphi_{1}$ or $\varphi_{2}$ where
appropriate. Furthermore, the flux now has the Eq.(\ref{eq:ExFlux})
form. To prevent ray dissipation by flux components normal to the
phase path, restraints must again be applied to the components for
$\mathbf{k}_{1}$ and $\mathbf{k}_{2}$. But these relations will
depend, in general, on the $\varphi_{1}$ and $\varphi_{2}$ values.
This therefore means that the flux is carried by phase points. These
localized energy bearers need to be marshaled into coordinated groups
by imposing some $\varphi_{1}$-$\varphi_{2}$ relation, such as $\varphi_{1}=\varphi_{2}=\varphi$,
to get classical fields.

The preceding component phase independence discussion simplicity can
be expected to be masked in real physical systems. For Hertz's dipole
field the electric field exhibits radial, transverse wave emergence
in the far field. At intermediate distances the waves have superluminal
radial speed that approaches light speed in the far field and the
waves change from longitudinal to transverse as the radial direction
changes from dipole length to dipole equator alignment. Hertz confirmed
these properties in the equatorial plane by observing interference
between free air and straight wire waves \cite[pp. 150-5]{Hertz}.

When phase point motion and wave flux are not aligned a flux component
normal to the phase point motion direction would force ray dissipation.
This is analogous to saying that all light rays are electromagnetic,
but not all electromagnetic waves are light rays. This analogy is
supported by synchrotron light emitted as rays by high speed electrons
in a circular orbit. These rays are attributed to the radial acceleration
not the periodic linear acceleration required to maintain the electron
energy, but they have not been identified in the electric field \cite{Synchrotron1,Synchrotron2,Synchrotron3}.
However, they are observed to be electrically polarized in the electron
orbit plane \cite{Elder}. These findings are expected if the rays
are phase directed polarization waves in the orbit plane. Although
component phase independence is mathematically compelling, its demonstrated
physical nonexistence would support treating polarizations as point
functions with closely bound potentials as described in Appendix \ref{sec:Polarization-Waves}
above. Even so, component phase independence is a classical field
hidden variable whose consequence is unintuitive. Whether electromagnetic
energy transport localization by phase independence is consistent
with quantum statistics will have to be examined separately. However,
the concept provides an unexplored means to help understand such light
generation problems as anisotropic emission from excited atom charge
distribution transitions for laser efficiency improvement.
\end{document}